# A Survey on Underwater Acoustic Sensor Networks: Perspectives on Protocol Design for Signaling, MAC and Routing

**Mohammad Sharif-Yazd[1], Mohammad Reza Khosravi[2], Mohammad Kazem Moghimi[3]\***

[1]Department of Electrical Engineering, Yazd Branch, Islamic Azad University, Yazd, Iran
[2]Department of Electrical and Electronic Engineering, Shiraz University of Technology, Shiraz, Iran
[3]Department of Electrical Engineering, Najafabad Branch, Islamic Azad University, Najafabad, Iran
Email: *moghimi.kazem@sel.iaun.ac.ir

## Abstract

Underwater acoustic sensor networks (UASNs) are often used for environmental and industrial sensing in undersea/ocean space, therefore, these networks are also named underwater wireless sensor networks (UWSNs). Underwater sensor networks are different from other sensor networks due to the acoustic channel used in their physical layer, thus we should discuss about the specific features of these underwater networks such as acoustic channel modeling and protocol design for different layers of open system interconnection (OSI) model. Each node of these networks as a sensor needs to exchange data with other nodes; however, complexity of the acoustic channel makes some challenges in practice, especially when we are designing the network protocols. Therefore based on the mentioned cases, we are going to review general issues of the design of a UASN in this paper. In this regard, we firstly describe the network architecture for a typical 3D UASN, then we review the characteristics of the acoustic channel and the corresponding challenges of it and finally, we discuss about the different layers e.g. MAC protocols, routing protocols, and signal processing for the application layer of UASNs.

## 1. Introduction

Monitoring of underwater environment is very important in marine science and technology. To cover this monitoring, creating underwater sensor networks is essential in undersea space. Besides, these networks have many sensor nodes with wireless links to each other. The wireless networks are used for underwater applications such as monitoring under the surface of seas and oceans and different applications in military, environmental and industrial usages. For example, there is need to underwater observatory to track submarines and to identify resources of pollution. Other applications can be noted are undersea monitoring for discovering the natural resources such as oil, gas and minerals [6]. To monitor these applications, it is necessary to have sensors in the underwater space. Since this monitoring is widely used, there are needs to be placed many of these sensors in there. Also to analyze the outputs of these sensors accurately, the sensors must be networked with each other in order to exchange information. Since establishment of a centralized network such as wired or access point-based wireless networks under the water surface especially in deep places is very expensive and in some cases is technically impossible or needs a long time, wireless networks in form of without structure (ad-hoc topology) are used. On the other hand in these new wireless networks, because the electromagnetic waves are very short-range, acoustic waves are used for communication between the sensors.

Figure 1 shows a sample 3D underwater sensor networks with hybrid architecture (some nodes are static and other nodes are mobile). Due to the high electromagnetic wave's attenuation in water, we must use acoustic waves instead of electromagnetic waves in underwater wireless networks. This issue makes new challenges related to the features of acoustic wave. The rest of this paper is as follows: at first, we review features of propagation model in acoustic channels, and then in the third section, we discuss the challenges of acoustic links in order to apply in underwater networking. Final section is allocated to conclusion.

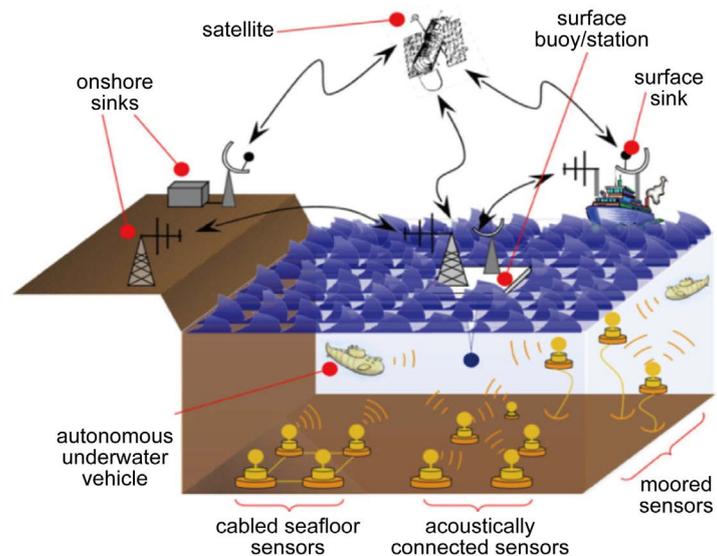

**Figure 1.** A typical 3D hybrid acoustic sensor network with static and mobile nodes; the network can be heterogeneous in terms of underwater space dimensions where consists of 2D and 3D sub-networks simultaneously, and additionally has mobile and static nodes (this figure has been extracted from [7]).

## 2. Propagation Model in Acoustic Channels

The specific features of the undersea environment affect energy consumption and traffic performance of the sensor nodes and also propagation of acoustic signal in the environment. The Thorp propagation model [1] is applied to explain the acoustic channel in underwater communications. As follows, we review some main details of the channel. The acoustic channel attenuation over distance $d$ can be expressed as below (Equation (1)) where $k$ denotes the geometric spreading ($k = 1.5$). In addition, $\alpha(f)$ represents the absorption coefficient.

$$A(d, f) = d^k \alpha(f)^d \tag{1}$$

The formulation of $\alpha(f)$ is expressed by the Thorp's propagation model as Equation (2), where $\alpha(f)$ is measured in dB/Km and $f$ is measured in KHz. The underwater environment noise can be expressed as Equation (3).

$$\alpha(f) = \begin{cases} \dfrac{0.11f^2}{1+f^2} + \dfrac{44f^2}{4100+f^2} + 2.75 \times 10^{-4} f^2 + 0.003 & f \geq 0.4 \\ 0.002 + 0.11\dfrac{f^2}{1+f^2} + 0.11f^2 & f < 0.4 \end{cases} \tag{2}$$

$$N(f) = N_t(f) + N_s(f) + N_w(f) + N_{th}(f) \tag{3}$$

In Equation (3), $N_t(f)$, $N_s(f)$, $N_w(f)$ and $N_{th}(f)$ denote different noises of the channel. $N_t(f)$ is caused by turbulence, $N_s(f)$ is by shipping movements, and also $N_w(f)$ and $N_{th}(f)$ are affected by waves and thermal/heat energy, respectively (see more details in [1]). For the acoustic signal with frequency of $f$ and propagation distance of $d$ in the underwater environment, signal to noise ratio at the receiver can be represented as the following equation.

$$SNR(f, d) = \dfrac{P(f)}{A(d, f)N(f)} \geq DT \tag{4}$$

In above equation, $P(f)$ shows the sending power at the sender. At the receiver, if $SNR(f, d)$ is not less than $DT$, the receiver can detect the received signals without error and correctly. $DT$ represents the detection threshold of the received signal. Therefore, we can dynamically set the transmission power of the sender according to the propagation distance, signal frequency and ambient control factors to decrease the consumed energy and prolong the network lifetime.

The underwater propagation velocity is a function of some things as temperature, pressure and salinity of seawater, which is written as Equation (5) [2], where $c$ shows the propagation velocity of sound in m/s; $T$ shows the temperature in degree Celsius; $S$ represents the salinity in parts per thousand, and $D$ is the depth in meters. It is observable that the sound velocity has direct relationship with temperature, depth and salinity. Equation (5) is valid when the conditions are satisfied as follows: $0 \leq T \leq 30$, $30 \leq S \leq 40$ and $0 \leq D \leq 8000$.

$$c = 1448.96 + 4.591T - 5.304 \times 0.01T^2 + 2.374 \times 0.01T^3 + 1.340(S - 35)$$
$$+ 1.63 \times 0.1D + 1.675 \times 10^{-7}D^2 - 1.025 \times 0.01T(S - 35) - 7.139 \times 10^{-13}TD^3 \quad (5)$$

## 3. UASNs' Challenges

In the previous section, we understood the specific features of an acoustic communication channel. These features create challenging issues in underwater sensor networks. In sub-sections of the current section, we review corresponding issues of different layers of UASNs from the PHY layer to application layer.

### 3.1. Physical (PHY) Layer

**Table 1** shows available bandwidth for different ranges in underwater acoustic channels, in fact, the features of underwater acoustic channel have caused these outputs in practical studies. Note that this table is independent from any signaling scheme (communication modulation, coding and etc.) and also transmission power, and in fact only shows available bandwidth in Hz. Explicitly, bandwidth is depended to distance from wave source that it is known as communication range in Km. Therefore, the bandwidth limitation is a major problem in the acoustic channels. **Table 2** shows the limited data rate in different signaling techniques as another major challenge which is directly related to the bandwidth limitation. It is noticeable that in the recent years, by use of efficient signaling in the limited bandwidth, data rate (bit per second) has increments in order of 10 Kbps. This table is only obtained according to limited transmission power that is operational and allowed for the underwater environment.

**Table 1.** Available bandwidth for different ranges in underwater acoustic channels [5].

| Coverage | Range (Km) | Bandwidth (KHz) |
|---|---|---|
| Very long | 1000 | Less than 1 |
| Long | 10 - 100 | 2 - 5 |
| Medium | 1 - 10 | 10 |
| Short | 0.1 - 1 | 20 - 50 |
| Very short | Less than 0.1 | More than 100 |

**Table 2.** Progress in digital modulation techniques for acoustic channels [4] [44]; notation (s) shows shallow waters and (d) shows deep waters.

| Type | Year | Rate (Kbps) | Range (Km) | Bandwidth (KHz) |
|---|---|---|---|---|
| FSK | 1984 | 1.2 | 3s | 5 |
| FSK | 1997 | 0.6 - 2.4 | 10d - 5s | 5 |
| DPSK | 1997 | 20 | 1d | 10 |
| QPSK | 1998 | 1.67 - 6.7 | 4d - 2s | 2 - 10 |
| 16-QAM | 2001 | 40 | 0.3s | 10 |
| 8-PSK | 2010 | 120 | 0.62d | 80 |

Another major problem in acoustic channels is high delay links in comparison to the electromagnetic channels which have carrier waves with speed similar to the light speed ($3 \times 10^8$ m/s). Speed of acoustic wave in underwater space is about 1500 m/s, but it is not fixed. This low speed creates high delay and it is very effective on the network performance. Complexity of acoustic channels versus electromagnetic channels is more than two issues of low bandwidth and high delay links, so main challenges have been listed in Table 3. This table also shows reasons of each challenge briefly. Some of them are main reasons of lower bandwidth of the acoustic links and the rest of them are main reasons for low data rate (Kbps) directly [6]. Bitrate is a main factor in traffic engineering of communication networks that is effective on quality of service (QoS) in exchanging multimedia contents, such as images and videos. For achieving a comprehensive view of key challenges of this layer, see Table 4.

### 3.2. MAC Protocols

Medium access control (MAC) is an important part of data link layer (the second layer in the OSI model). Main duty of MAC is channel allocation or in the other words, bandwidth multiplexing based on multiple access techniques in networks. Hereby, MAC protocols in addition to modulation and channel coding techniques which are used in PHY layer, can be known a way for enhancing the network signaling. Of course, there are some appearances of signaling in the application layer, *i.e.*, control of the network performance based on application layer tools and efficient signal processing techniques. Generally in wireless net-works, MAC protocols are classified into two categories, contention-based techniques (random access techniques or techniques with shared channel, e.g. ALOHA, carrier sense multiple access (CSMA) and its applied versions) and channelization techniques (deterministic techniques with unshared channel, e.g. frequency division multiple access (FDMA), time division multiple access (TDMA), code division multiple access (CDMA [42]) and space division multiple access (SDMA)) [8] [9]. The low bandwidth issue of the underwater acoustic links causes that in practice, the second category often becomes unsuitable in these links, however, there are some specific cases based on channelization techniques, see more about them in [10] [11]. Assume that number of the sensor nodes is N, then for creating an unshared channel, we have to create too large number of independent connections between all sensor nodes, therefore, the number of connections for all the network can be observed in Equation (6). When N becomes greater than about 50 nodes (under each volume of 3D space; in practice number of nodes in comparison to volume of 3D space affects the density of a network; where the density is depended on some variables such as number of nodes, sensing coverage, communication coverage, volume of space and maybe mobility of nodes in some cases), consequently, we need at least 1250 independent sub-channels. Therefore it is noticeable that in dense underwater networks, MAC protocols with the shared channel are more proper than the unshared channels.

**Table 3.** A brief list of the challenges in PHY layer based on the acoustic channel.

| Challenges | Reason(s) | Effect(s) |
|---|---|---|
| Path loss | Geometric spreading/Energy loss | Low bandwidth/Low data rate |
| Shadowing | Random distribution of objects in 3D space | Low bandwidth/Low data rate |
| Noisy channel | Human made noise/Ambient noise/Other noises | Low data rate (Hard detection) |
| Hard channel fading | Multipath | Low data rate (Hard detection) |
| Inter-Symbol-Interference (ISI) | High communication rate in low bandwidth | Low data rate (Hard detection) |
| Non-fixed delay | Non-fixed speed of carrier wave | Low data rate (Hard detection) |
| Doppler shift | Non-fixed frequency of carrier wave | Low data rate (Hard detection) |

**Table 4.** Main concluded challenges of PHY layer.

| Effects | Intrinsic factor (Directly) | Controllable factor | Solutions |
|---|---|---|---|
| Low bandwidth | Yes | No | Only by control of data rate |
| Low data rate | No | Yes | Efficient and high performance signaling |
| High delay | Yes (Propagation delay) | Yes (Processing delay) | Reduction of processing delay |

$$\text{Number of Independent Connections} \left( \text{NIC} \right) = \binom{N}{2} = \frac{N \left( N - 1 \right)}{2}$$

$$\overset{\text{If } N = 50}{\rightarrow} \quad \text{NIC} = 1250 \tag{6}$$

$$\overset{\text{If Bitrate} = 100 \text{ Kbps}}{\rightarrow} \quad \text{Bitrate of each channel} = \frac{100 \text{ Kbps}}{1250} < 100 \text{ bps}$$

Some of the most famous MAC protocols for the acoustic channels which all of them have often been derived based on CSMA (they are enhanced versions of CSMA with collision avoidance (CSMA/CA)), are underwater MAC/broadcasting [6], T-MAC and Tu-MAC [12] and so on.

### 3.3. Routing Protocols

Classically, design of routing protocols for terrestrial (electromagnetic) wireless sensor networks is according to path-based policy, namely, routing process selects a path with lowest cost for packet forwarding based on a computational criterion. Therefore, this type of routing protocols is named single-path routing. However, high delay of the acoustic links concurrently with mobility of the sensor nodes makes path-based routing techniques inefficient, because under any cost function, while a path (shortest path) is selected for forwarding process, variation of the network topology makes the selected path invalid very soon. Basically, three types of mobility exist in a WSN, static topology, low-speed topology and high-speed topology where two last cases are dynamic; especially when WSNs are used for underwater environment, their topology is mainly dy-namic, whether variation of topology is low-speed/smooth (only by external force of marine streams) or high-speed/fast (with both external and internal forces). Consequently in the recent years, design of underwater routing protocols is often based on the flooding algorithm [9]. Nowadays, flooding-based routing is widely used, although it has an intrinsic inefficiency due to the use of multi-path process. However in practice, the multi-path techniques are better than the single-path techniques in the most cases (specifically in the high-speed topology) in terms of terrific and even lifetime performance. Of course, there are some path-based protocols in this respect based on TDMA, OFDMA and so on, read more information about them in [13] [45]. Two most famous multi-path routing protocols are VBF [17] [41] and DBR [18]. Already, there are many extensions based on both, for example EE-DBR [14], D-DBR [15], EELD-DBR [16] for DBR and HH-VBF [19], AHH-VBF [3], CDZ-VBF [20], SD-VBF [6] and RC-VBF [21] for VBF. **Figure 2** shows the details of basic VBF protocol as a multi-path technique. In multi-path techniques, there are several paths between the sender and the receiver which they create a redundancy for the problem of path loss in the underwater channel.

### 3.4. Transport Layer

Classic protocols of this layer are transmission control protocol (TCP) and user datagram protocol (UDP) [8] [9]. High delay of the links in UASNs is compatible with UDP, because UDP has lower processing delay than TCP, therefore it can obtain the metrics of quality of service (QoS) efficiently. A research topic for this layer is the topology control problem of wireless networks.

### 3.5. Application Layer and Signal Processing

Application layer and signal processing issues in WSNs and UASNs are so hot and new topics for the related researches. Previously, application layer of WSNs has been considered similar to the internet network and according to TCP/IP standard (at most). However, we newly see that some studies on the application layer of WSNs are done which propose novel strategies for the specific usages of WSNs. These researches pay attention to some traffic tools of the application layer, security of the layer and signal processing algorithms related to the layer. Combination of WSNs and new topics such as internet of things, big data, and software defined networks (SDNs) are currently hot areas for researches. However, developing the mentioned cases in UASNs is completely in first steps and number of researches is limited (for example [43]), therefore, it is a good area for future researches. According to the technical progress of electronic chips in the recent years, the sensor nodes of an UASN can do data processing more than the past decades. In Figure 3, part (a) shows the integration and processing of all received data from all sensor nodes of a WSN by a final central processing unit (CPU) at the receiver side, where in this architecture, the sensor nodes are not able to do the complete processing. Part (b) shows the elements of a smart sensor which has a single-chip microprocessor, where in this second architecture, all processes on the sensed data of all sensors can be performed by themselves; in the other words, WSN operates as a distributed system with parallel processing. Some of signal processing algorithms which have already been proposed for WSNs are mainly about visual sensor networks and video communications over WSNs [23] [24] [25] [26] [27]. Therefore, the most of aspects in signal processing can be adapted with the applications related to UASNs, for example information fusion, image and video processing [28]-[40].

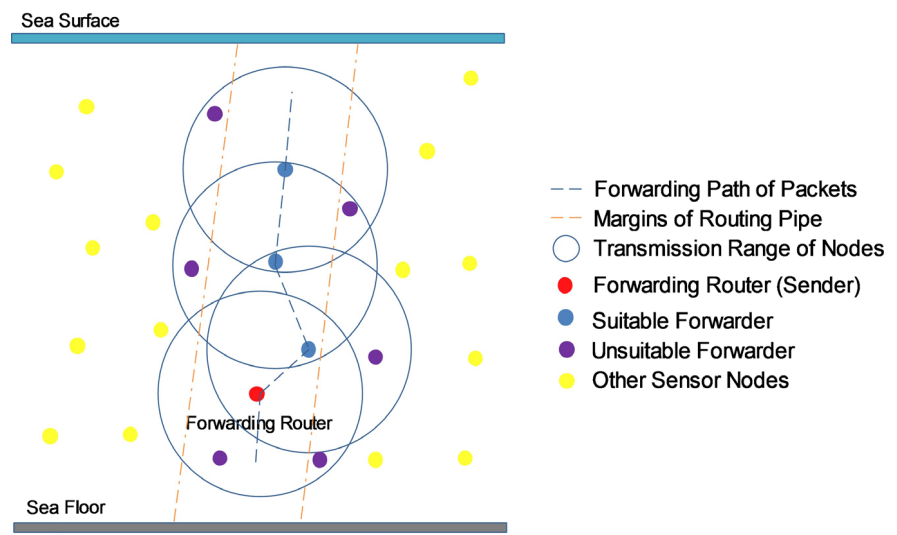

**Figure 2.** Schematic of VBF.

### 3.6. Integrated Challenges and Cross-Layer Constraints

As a final review, we integrate all challenges and constraints of UASNs in Figure 4. Attention to all aspects of these challenges is very important for design of an UASN. For more details and description in this respect, refer to [12].

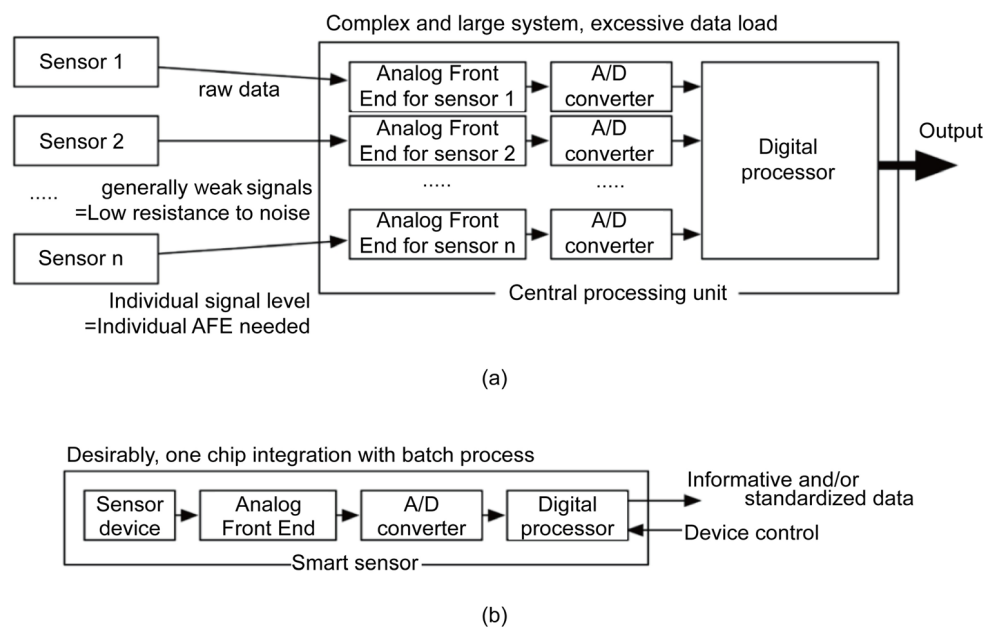

(a)

(b)

**Figure 3.** Signal processing unit for WSNs [22] can also be used in UASNs. Parts (a) and (b) are related to systems with centralized and distributed processing, respectively.

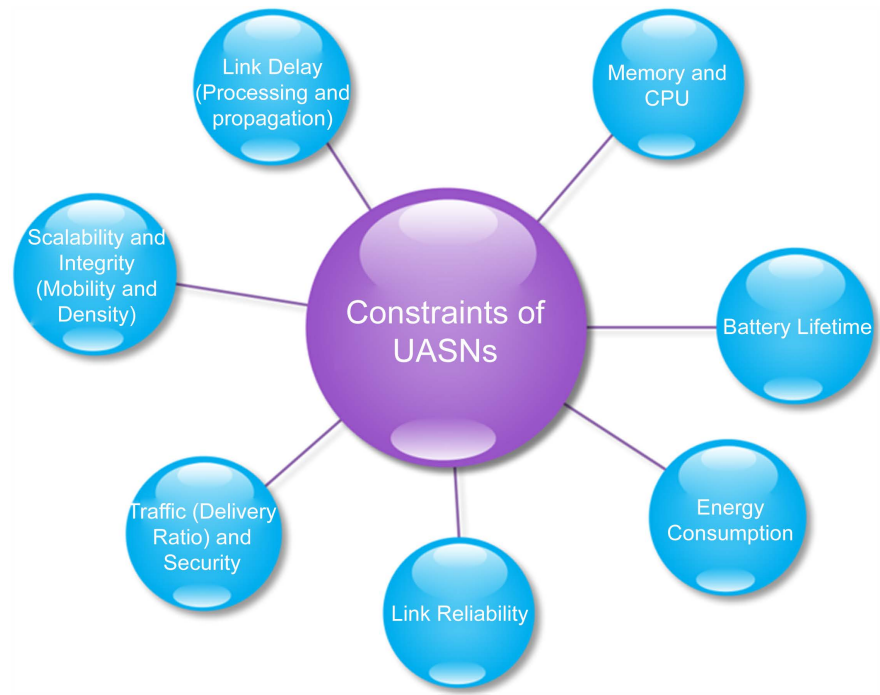

**Figure 4.** A general overview on UASNs' constraints for designing the network protocols; percentage of importance of these mentioned cases in different applications may be various; however, the most important cases are often energy consumption and traffic performance.

## 4. Conclusion

In this paper, we reviewed the features and challenges of the underwater acoustic links in order to apply in underwater sensor networks. Generally, main idea of this survey is creation of a perception of underwater acoustic sensor networks' challenges. The discussed items contained PHY layer constraint, MAC and routing design, and new topics regarding signal processing of UASNs. We find out that all of these challenges create networking complexity; therefore, attention to these challenges is an essential condition while designing the efficient network protocols.